\documentclass[reprint, amsmath, amssymb, aps, prd, superscriptaddress, nofootinbib, nobibnotes, longbibliography, floatfix, colorlinks=true, linkcolor=red, citecolor=blue]{revtex4-2}

\usepackage{graphicx}
\usepackage{float}
\graphicspath{{./figures/}}
\usepackage{orcidlink}

\usepackage{amsmath}
\usepackage{amsfonts}
\usepackage{amssymb}
\usepackage{multirow}
\usepackage{hyperref}

\usepackage[capitalise]{cleveref}
\crefname{figure}{Fig.}{Figs.}
\Crefname{figure}{Fig.}{Figs.}

\def\({\left(}
\def\){\right)}
\def\[{\left[}
\def\]{\right]}

\def\be{\begin{equation}}
\def\ee{\end{equation}}
\def\beq{\begin{eqnarray}}
\def\eeq{\end{eqnarray}}

\usepackage{acro}
\DeclareAcronym{BH}{
	short = BH,
	long  = black hole
}

\DeclareAcronym{GR}{
	short = GR,
	long  = general relativity
	}

\DeclareAcronym{BBH}{
	short = BBH,
	long  = binary black hole
}
\DeclareAcronym{GW}{
	short = GW,
	long  = gravitational wave
}
\DeclareAcronym{BNS}{
	short = BNS,
	long  = binary neutron star
}
\DeclareAcronym{CBC}{
	short = CBC,
	long  = compact binary coalescence
}
\DeclareAcronym{SNR}{
	short = SNR,
	long  = signal-to-noise ratio
}
\DeclareAcronym{KDE}{
	short = KDE,
	long  = kernel density estimation
}

\DeclareAcronym{GPR}{
	short = GPR,
	long  = Gaussian process regression
}

\DeclareAcronym{TIGER}{
	short = TIGER,
	long  = Test Infrastructure for General Relativity
}

\DeclareAcronym{MI}{
	short = MI,
	long  = mutual information
}

\begin{document}

%\begin{CJK*}{UTF8}{gbsn}
\title{
   Testing Source Dependence in Gravitational-Wave Tests of General Relativity with Mutual Information
}
\author{Yi-Fan Wang \orcidlink{0000-0002-2928-2916}}
\email{yifan.wang@aei.mpg.de}
\affiliation{Max-Planck-Institut f{\"u}r Gravitationsphysik (Albert-Einstein-Institut), Am M{\"u}hlenberg 1, D-14476 Potsdam, Germany}
\affiliation{Purple Mountain Observatory, Chinese Academy of Sciences, Nanjing 210034, China}
\begin{abstract}
Phenomenological deviation parameters used in gravitational-wave tests of general relativity need not represent universal fundamental properties of the sources. Instead, a deviation parameter can be a derived quantity that depends on source properties. Apparent violations of general relativity arising from waveform systematics or data-quality issues can also depend on source properties. In addition to diagnosing individual events, it is helpful to identify systematic trends at the population level. We introduce a nonparametric framework that uses mutual information to test whether the deviations depend on source properties. We validate the framework using simulations with massive graviton waveforms. Assuming we do not have prior knowledge of the underlying theory, massive graviton effects can be captured by a parameterized post-Einstein coefficient that is expected to depend on the luminosity distance and source mass. The results show that the mass and distance dependencies become clear for graviton masses of $\geq 1\times10^{-22}\,\mathrm{eV}$ at detector sensitivities representative of the current fourth observing run, while the controlled general relativity simulations yield null results. We then analyze public real-data products for tests of general relativity from GWTC-4.0 and 4-OGC, including phenomenological parameterized tests, modified dispersion tests, and parity violation tests. The phenomenological deviation parameters have tentative correlations with effective spin, and correlations of modified-dispersion or parity-violation parameters with source mass are visible. These results demonstrate that catalog-level dependence can identify patterns and thus be used to diagnose the origin of deviations in testing general relativity. The framework therefore complements single-event analyses and parameterized hierarchical population tests.
\end{abstract} 
%\end{CJK*}

\maketitle
\section{Introduction}
Observations of gravitational waves from compact binary coalescences have enabled precise tests of \ac{GR} in the strong-field and dynamical regime ever since the first event GW150914 \cite{GW150914}.
For that event, Ref.~\cite{GW150914testingGR} performed theory-agnostic parameterized waveform tests, inspiral--merger--ringdown consistency tests, ringdown analyses, and a constraint on the graviton mass.
Subsequent observations of more black hole events further tightened the constraints and added new tests, such as those of Lorentz-violating dispersion~\cite{LIGOScientific:2019fpa}.
The binary neutron star merger GW170817 \cite{GW170817} enabled tests of the speed of gravitational waves and extra dimensions, among others~\cite{LIGOScientific:2018dkp}.
With the rapidly growing number of observed events with diverse properties, these tests now cover various aspects of gravitational waves, such as their generation, propagation, and remnant black hole physics~\cite{LIGOScientific:2026overview,LIGOScientific:2026fcf,LIGOScientific:2026remnants,LIGOScientific:2020tif,LIGOScientific:2021sio}.
So far, these studies have found no compelling evidence for a violation of \ac{GR} \cite{Krishnendu:2021fga,Yunes:2025xwp,livingreviewWill}.
Nevertheless, the enlarged catalog has tightened constraints on deviations from individual events and created an opportunity to search for systematic trends that cannot be recognized from a single event alone.

Many tests of \ac{GR} introduce phenomenological deviation parameters without committing to a particular alternative theory.
These parameters describe departures from different aspects of a \ac{GR} waveform, and \ac{GR} is usually recovered when they vanish.
For example, the parameterized post-Einstein (ppE) waveform framework~\cite{Yunes:2009ke} introduces phenomenological deviations in the post-Newtonian coefficients of the inspiral phase and calibration parameters of the merger and ringdown phases.
Bayesian analyses are then used to infer the parameters event by event and compare the Bayesian evidence for a beyond-\ac{GR} hypothesis with \ac{GR}~\cite{Li:2011cg}.
Combining multiple events then requires an explicit assumption about how each deviation parameter behaves across the catalog.
A combination may assume a common value, treat events independently, or model the values as draws from a parameterized population distribution, for example, a Gaussian distribution~\cite{Isi:2019asy}.

This approach is useful because the same analysis can be mapped onto many possible departures from \ac{GR}. However, this generality also limits its interpretation: a phenomenological coefficient need not be a fundamental quantity with the same value across different events or follow an assumed population distribution.
Instead, the deviation may depend on physical properties such as source mass or distance.
Even if \ac{GR} is correct, an apparent deviation may occur due to waveform inaccuracies \cite{Chandramouli:2024vhw,Saini:2022igm,Moore:2021eok,SciPostPhysCommRep.5}, detector or data-quality effects \cite{Payne:2022spz,Udall:2024ovp}, or limitations of the chosen parameterization and priors, as discussed, for example, in Refs.~\cite{Payne:2023kwj, Sanger:2024axs}.
In such a case, the deviation may also depend on a source property or detector status, such as a region where the waveform model is less accurate, or a particular time when the detector is not well calibrated.

This motivates a question complementary to conventional single-event tests of \ac{GR} and parameterized hierarchical population tests \cite{Isi:2019asy, Isi:2022cii,Payne:2023kwj,Zimmerman:2019wzo}: do the inferred deviations depend on the physical properties of the sources?
Candidate properties include total masses, mass ratio, spins, eccentricity, anomaly, orientation, luminosity distance, sky location, \ac{SNR}, etc.
Such dependence contains information that can reveal the origin of the apparent violation.
Past work addressing this question includes \cite{Dideron:2022tap, Dideron:2024xwm,Dideron:2026uis} that uses a structured correlated residual method to search for unmodeled physics and the dependence on mass.
This information is lost when a catalog is represented only by a common deviation parameter.
An example of this kind is a massive graviton~\cite{Will:1997bb,Mirshekari:2011yq}.
A nonzero graviton mass introduces a frequency-dependent phase correction that accumulates during propagation.
Assuming we do not have prior knowledge of massive graviton theory and test \ac{GR} with a phenomenological parameterized post-Einstein waveform model, the inferred deviation parameter is expected to depend on both propagation distance and source mass.
Thus, detecting this dependence would reveal crucial information about the underlying theory.
Another example arises from apparent violations.
A waveform model that neglects higher-order modes can bias testing \ac{GR} parameters preferentially for asymmetric binaries or particular viewing orientations, as shown in Ref.~\cite{Pang:2018hjb}.
Therefore, a dependence on mass ratio and inclination would appear in such a case, and detecting such a correlation would indicate waveform systematics.

We use \ac{MI} to search for general dependencies between the inferred deviation parameters and source properties.
\ac{MI} is an information-theoretic measure that identifies general statistical dependence, not merely linear correlation.
It does not assume a predetermined functional form.
We estimate it from the full event-level posteriors so that each event retains its measurement uncertainty and the within-event covariance between the deviation parameter and source properties. 
We use a permutation-based method to estimate its null distribution and hence its statistical significance.
The framework is validated with an injection and recovery campaign using massive graviton waveforms.
We generate four 100-event catalogs with $m_g=\{0,1,3,5\}\times10^{-22}\,\mathrm{eV}$ using the same sources, detector configuration, and noise realizations. Each catalog is recovered with a parameterized post-Einstein coefficient.
The dependence of the recovered coefficient on source chirp mass and distance is strongest for the two largest injected graviton masses, in the direction predicted by the underlying theory. It is also visible for $m_g=1\times10^{-22}\,\mathrm{eV}$, while the controlled \ac{GR} catalog shows no significant correlation.
We then apply the same analysis to the public 4-OGC and GWTC-4.0 testing-\ac{GR} products \cite{ligo_scientific_collaboration_2026_21403342}, including phenomenological parameterized tests (TIGER and FTI; see \cref{sec:real-data}), modified-dispersion tests, and parity-violation tests.
We find that the TIGER and FTI tests have tentative correlations with effective spin, while the modified-dispersion and parity-violation tests show visible dependence on mass.
These observations are used to diagnose the origin of these apparent deviations.

The remainder of this paper is organized as follows.
In \cref{sec:TIGER}, we review the phenomenological parameterized testing-\ac{GR} framework and the interpretation of event-level deviation parameters.
\Cref{sec:method} introduces mutual information, conditional mutual information, posterior resampling, and event permutation based $p$-value estimation.
In \cref{sec:massive-graviton-sim}, we validate the method using the massive graviton injection and recovery study.
\Cref{sec:real-data} presents applications to public gravitational-wave testing-\ac{GR} results.
We discuss the physical interpretation and possible systematic origins of the observed dependence structures in \cref{sec:discussion-conclusion}.

\section{Parameter estimation for parameterized tests of general relativity}
\label{sec:TIGER}

\subsection{Single-event Bayesian inference}

For a gravitational-wave event, the detector data $d$ are analyzed by comparing them with a waveform model $h(\boldsymbol{\theta})$.
We separate the waveform parameters into two groups,
\begin{equation}
\boldsymbol{\theta}=\left(\boldsymbol{\lambda},\boldsymbol{\delta}\right),
\end{equation}
where $\boldsymbol{\lambda}$ denotes the ordinary source parameters within the prediction of \ac{GR}, and $\boldsymbol{\delta}$ denotes one or more phenomenological \ac{GR} deviation parameters.
The former may include the component masses, spins, eccentricity and its anomaly, luminosity distance, sky position, binary orientation, polarization, and coalescence time.
The latter quantify departures from a \ac{GR} waveform model.

Within a model $\mathcal{H}$, Bayesian inference gives the joint posterior
\begin{equation}
p\left(\boldsymbol{\lambda},\boldsymbol{\delta}\mid d,\mathcal{H}\right) =\frac{p\left(d\mid\boldsymbol{\lambda},\boldsymbol{\delta},\mathcal{H}\right)\pi\left(\boldsymbol{\lambda},\boldsymbol{\delta}\mid\mathcal{H}\right)}{\mathcal{Z}(d\mid\mathcal{H})},
\label{eq:event-posterior}
\end{equation}
where $\pi$ is the prior and $\mathcal{Z}$ is the Bayesian evidence.
For stationary Gaussian detector noise, the likelihood is proportional to
\begin{equation}
p\left(d\mid\boldsymbol{\lambda},\boldsymbol{\delta},\mathcal{H}\right) \propto\exp\left[-\frac{1}{2}\sum_D\left\langle r^D\mid r^D\right\rangle_D\right],
\label{eq:gw-likelihood}
\end{equation}
where the sum runs over detectors and
\begin{equation}
r^D=d^D-h^D\left(\boldsymbol{\lambda},\boldsymbol{\delta}\right),
\end{equation}
is the residual in detector $D$. The noise-weighted inner product is
\begin{equation}
\left\langle a\mid b\right\rangle_D=4\,\Re\int_{f_{\mathrm{low}}}^{f_{\mathrm{high}}}\frac{\tilde{a}(f)\tilde{b}^{*}(f)}{S_{n,D}(f)}\,\mathrm{d}f,
\label{eq:noise-inner-product}
\end{equation}
with $S_{n,D}(f)$ the one-sided noise power spectral density.

\Cref{eq:event-posterior} is the basic product provided by a parameterized testing-\ac{GR} analysis and is used in this work.
In particular, we require posterior information for both a phenomenological deviation parameter and the candidate source properties.

\subsection{Phenomenological parameterizations of deviations from general relativity}

Theory-agnostic tests introduce parameterized deformations of a \ac{GR} waveform without committing to a particular alternative theory~\cite{Li:2011cg,Agathos:2013upa,LIGOScientific:2026fcf}.
Schematically, a frequency-domain waveform can be written as
\begin{equation}
\tilde{h}(f)=\mathcal{A}\left(f;\boldsymbol{\lambda},\boldsymbol{\delta}\right)\exp\left[i\Psi\left(f;\boldsymbol{\lambda},\boldsymbol{\delta}\right)\right],
\label{eq:deformed-waveform}
\end{equation}
where the \ac{GR} waveform is typically recovered at $\boldsymbol{\delta}=\boldsymbol{0}$.
The parameterized post-Einstein framework is established as follows.
Taking the inspiral of a binary black hole as an example, the Fourier-domain phase can be expanded systematically as
\begin{equation}
\Psi_{\mathrm{GR}}(f)=2\pi f t_c-\phi_c-\frac{\pi}{4}+\frac{3}{128\eta v^5}\sum_k\left[\varphi_k+\varphi_k^{\log}\ln v\right]v^k,
\label{eq:pn-phase}
\end{equation}
where $v=(\pi M f)^{1/3}$, $M$ is the total mass, $\eta$ is the symmetric mass ratio, and $t_c$ and $\phi_c$ are the coalescence time and phase.
The post-Newtonian coefficients $\varphi_k$ and $\varphi_k^{\log}$ are determined as functions of the intrinsic source parameters in \ac{GR}~\cite{Buonanno:2009zt}.

Generically, a phenomenological phase deformation can be added as
\begin{equation}
\Psi(f)=\Psi_{\mathrm{GR}}(f)+\sum_i\delta_i B_i\left(f;\boldsymbol{\lambda}\right),
\label{eq:generic-phase-deformation}
\end{equation}
where $B_i$ specifies the frequency dependence of the deformation and $\delta_i$ is the deformation parameter.
For the inspiral parameterized post-Einstein convention used in this work, these basis functions and coefficients are
\begin{equation}
\Delta\Psi_{\rm ppE}(f) = \sum_i\beta_i\left(\pi\mathcal{M}_{\rm det}f\right)^{(i-5)/3},
\label{eq:ppe-basis-function}
\end{equation}
where $\mathcal{M}_{\rm det}$ is the detector-frame chirp mass and $\beta_i$ is the phenomenological coefficient associated with the $f^{(i-5)/3}$ basis.
In particular, $i=2$ gives $B_2(f)=1/(\pi\mathcal{M}_{\rm det}f)$ with the $f^{-1}$ term corresponding to the massive graviton correction (see \cref{sec:massive-graviton-sim}).
The same type of modification can also be applied to the intermediate and merger--ringdown parts of a waveform~\cite{IMRPhenomD1,IMRPhenomD2}.

This notation reflects the phenomenological feature that, although $\delta_i$ is varied as an additional inference parameter, its relation to a more fundamental universal constant may depend on source parameters $\boldsymbol{\lambda}$.
The aim of this work is to identify whether such a dependence exists in the observed catalog.

\subsection{From event-level posteriors to a catalog-level dependence test}

For a catalog of $N$ detected events labeled by the index $i$, the direct outputs of parameterized tests of \ac{GR} are the event-level posteriors $p_i(\boldsymbol{\delta}_i,\boldsymbol{\lambda}_i \mid d_i,\mathcal{H})$.
To formulate the dependence test, let $X_i$ denote a draw from $\boldsymbol{\delta}_i$, and let $Y_i$ be a paired draw from a source or measurement property.
The latter can be a source parameter associated with $\boldsymbol{\lambda}_i$, such as mass or spin, or other properties such as \ac{SNR}.
We denote by $p_i^{XY}(x,y)$ the corresponding event-level joint posterior and $q_{\mathrm{cat}}$ the target catalog-level distribution for $(x,y)$.
Note that representing $q_{\mathrm{cat}}$ as a mixture of the event-level posteriors
requires a choice of event weights.
We assume equal weights for different events in this work
\begin{equation}
q_{\mathrm{cat}}(x,y \mid \{d_i\}_{i=1}^{N},\mathcal{H})
=
\frac{1}{N}\sum_{i=1}^{N}p_i^{XY}(x,y).
\label{eq:catalog-posterior-mixture}
\end{equation}
Equal event weighting is just a simplified assumption when the underlying population of \ac{GR} violations is unknown.
More generally, the factor $1/N$ can be replaced by weights $w_i$ satisfying $\sum_{i=1}^{N}w_i=1$ \cite{Mandel:2018mve}. Such weights can account for selection effects, which may change the observed dependence structure.
Interpreting a detected association as a property of the intrinsic population requires a generative population model and end-to-end simulations that account for selection effects.
For an explicit model of selection effects from an intrinsic population model, see Ref.~\cite{Magee:2023muf}.

Our null hypothesis is that the deviation parameter and the event property are statistically independent in this catalog,
\begin{equation}
H_0:\quad
q_{\mathrm{cat}}(x,y) = q_{\mathrm{cat}}^{X}(x)\,q_{\mathrm{cat}}^{Y}(y),
\label{eq:independence-hypothesis}
\end{equation}
whereas the alternative is
\begin{equation}
H_1:\quad
q_{\mathrm{cat}}(x,y) \neq q_{\mathrm{cat}}^{X}(x)\, q_{\mathrm{cat}}^{Y}(y).
\label{eq:dependence-hypothesis}
\end{equation}
No functional form is specified under $H_1$.
As introduced in the next section, \ac{MI} is used as a test statistic, and an event permutation based null test is used to quantify the significance of rejecting $H_0$.

As a special case, the null hypothesis is retained when all $\delta=0$.
When a fraction of events show nonzero deviations, $H_1$ may be preferred if those deviations have a systematic trend with $Y$.
This is physically plausible if an underlying modified gravity is characterized by a coupling $\kappa$, and its effective phenomenological deviation parameter takes the form
\begin{equation}
X_i = F\left(\kappa,Y_i\right).
\label{eq:physical-to-phenomenological-map}
\end{equation}
A common value of $\kappa$ can therefore produce different $X$ for sources with different $Y_i$.
The $H_0$ hypothesis can also fail when waveform inaccuracies, prior effects, detector effects, or measurement precision vary across the source-parameter space.
Rejecting $H_0$ consequently reveals a systematic trend, thus providing hints for the physical origin of the apparent violation of \ac{GR}.

This test is complementary to the hierarchical combination of multiple events with a parameterized distribution model.
Such a model may assume a universal value, $X_i=X_\star$.
Another model may instead assume a Gaussian distribution of $X_i$ across the catalog, that is,
\begin{equation}
X_i\sim\mathcal{N}(\mu,\sigma^2),
\end{equation}
which is independent of the source properties~\cite{Isi:2019asy}.
Ref.~\cite{Payne:2024yhk} has also considered a parameterized relation between the deviation parameter and source mass.

\section{Mutual information and catalog-level dependence tests}
\label{sec:method}

\subsection{Mutual information}

We quantify the departure from the independence hypothesis $H_0$ in Eq.~\eqref{eq:independence-hypothesis} using \ac{MI}.
In the notations of this work, the \ac{MI} between a deviation parameter $X$ and an event property $Y$ is defined as~\cite{Shannon}
\begin{equation}
I(X;Y)=\int q_{\mathrm{cat}}(x,y)\ln\!\left[\frac{q_{\mathrm{cat}}(x,y)}{q_{\mathrm{cat}}^{X}(x)q_{\mathrm{cat}}^{Y}(y)}\right]\,\mathrm{d}x\,\mathrm{d}y.
\label{eq:mutual-information}
\end{equation}
This is also the Kullback--Leibler (KL) divergence between the joint distribution $q_{\mathrm{cat}}(x,y)$ and the product of its marginals $q_{\mathrm{cat}}^{X}(x)$ and $q_{\mathrm{cat}}^{Y}(y)$.
It is non-negative, measured in nats, and vanishes if and only if $X$ and $Y$ are statistically independent.
Unlike a linear correlation coefficient, \ac{MI} can identify any dependence, including nonlinear and nonmonotonic patterns, without committing to a functional form.
Its value quantifies the strength of correlation in the analyzed catalog.
\ac{MI} is infinite when $Y$ is a function of $X$.

When considering a third property $Z$, we also use conditional mutual information (CMI), which is defined as the expectation of the \ac{MI} between $X$ and $Y$ conditioned on $Z$:
\begin{align}
I(X;Y\mid Z)
&=\int q_{\mathrm{cat}}(z)q_{\mathrm{cat}}(x,y\mid z)\nonumber\\
&\qquad\times\ln\!\left[\frac{q_{\mathrm{cat}}(x,y\mid z)}{q_{\mathrm{cat}}(x\mid z)q_{\mathrm{cat}}(y\mid z)}\right]
\,\mathrm{d}x\,\mathrm{d}y\,\mathrm{d}z.
\label{eq:conditional-mutual-information}
\end{align}
CMI measures the residual dependence between $X$ and $Y$ after conditioning on $Z$ and vanishes when they are conditionally independent given $Z$.
We use it to ask, for example, whether a distance dependence remains after conditioning on source mass in the massive graviton example.
The variables $X$, $Y$, and $Z$ may also be multidimensional. For example, $Y$ may represent a two-dimensional sky location in a test of anisotropic modifications of gravity.

\subsection{Estimation of mutual information from posterior samples}

The event-level posteriors are typically represented by numerical samples rather than analytic distributions, so Eq.~\eqref{eq:mutual-information} must be estimated from posterior samples.
To compute \ac{MI}, we repeatedly construct a catalog realization.
For each realization $r$, we draw one sample from each event,
\begin{equation}
\left(X_i^{(r)},Y_i^{(r)}\right)\sim p_i^{XY},\qquad i=1,\ldots,N,
\label{eq:posterior-resampled-catalog}
\end{equation}
where $p_i^{XY}$ is the joint marginal posterior of the selected pair of variables for event $i$.
These draws form the posterior-resampled catalog $\mathcal{C}^{(r)}=\{(X_i^{(r)},Y_i^{(r)})\}_{i=1}^{N}$ by repeating $R$ times.
For CMI, the same construction includes $Z_i^{(r)}$, with $(X_i^{(r)},Y_i^{(r)},Z_i^{(r)})$ drawn from the corresponding joint three-variable posterior.
In this work, we use $R=2000$.
This procedure naturally propagates the measurement uncertainty of each event into the distribution of estimated \ac{MI} values.
Because every realization contains exactly one draw from each event, the procedure implements the equal event weighting assumed in Eq.~\eqref{eq:catalog-posterior-mixture}.
%Whenever joint posterior samples are available, $X$ and $Y$ are taken from the same posterior, preserving their within-event covariance. For some data products, however, only one-dimensional testing-\ac{GR} marginals are available, while the source-parameter posteriors are released separately in standard GWTC analyses within \ac{GR}. We then have to sample the variables independently; however, this procedure does not preserve the within-event covariance and is stated explicitly wherever it is used.

Before estimating \ac{MI}, we standardize the variables because their different numerical scales affect the nearest-neighbour distance estimation (see below).
For a source-property variable $Y$, we subtract its mean $\mu_Y$ and divide by the standard deviation $\sigma_Y$.
For a deviation variable $X$, we consider its mean to zero to measure its deviation from the \ac{GR} reference value. Thus,
\begin{equation}
\widetilde{X}_i^{(r)}=\frac{X_i^{(r)}}{\sigma_X},\qquad
\widetilde{Y}_{i}^{(r)}=\frac{Y_{i}^{(r)}-\mu_Y}{\sigma_Y},
\label{eq:mi-standardization}
\end{equation}
where $\sigma_{X/Y}$ and $\mu_{X/Y}$ are computed as the ensemble average from each resampled realization.
$Z$ is standardized the same way when it is included.
The \ac{MI} is invariant under these affine transformations.

We use the Kraskov--St\"ogbauer--Grassberger (KSG) $k$-nearest-neighbour estimator~\cite{MIPRE} to compute \ac{MI}.
For each event $i$, let $\epsilon_i$ be the distance to its $k$-th nearest neighbour in the joint $(X,Y)$ space under the maximum norm, and let $n_x(i)$ and $n_y(i)$ count the number of other events strictly within the same radius in the two marginal spaces.
The estimator is
\begin{align}
\widehat{I}(X;Y)
&=\psi(k)+\psi(N)-\frac{1}{N}\sum_{i=1}^{N}\left[\psi\!\left(n_x(i)+1\right)\right.\nonumber\\
&\hspace{42mm}\left.+\psi\!\left(n_y(i)+1\right)\right],
\label{eq:ksg-estimator}
\end{align}
where $\psi$ is the digamma function.
We set $k=3$ and clip negative estimates of \ac{MI}, which can occur because of finite-sample effects, to zero \cite{PhysRevE.100.022404}.

We estimate CMI with the corresponding Frenzel--Pompe nearest-neighbour estimator~\cite{FrenzelPompe2007}.
For each event $i$, the radius $\epsilon_i$ is set by its $k$-th neighbour in the joint $(X,Y,Z)$ space.
Let $n_z(i)$, $n_{xz}(i)$, and $n_{yz}(i)$ denote the numbers of other events strictly within this radius in $Z$, $(X,Z)$, and $(Y,Z)$, respectively.
The estimator is
\begin{align}
\widehat{I}(X;Y\mid Z)
&=\psi(k)+\frac{1}{N}\sum_{i=1}^{N}\left[\psi\!\left(n_z(i)+1\right)\right.\nonumber\\
&\qquad\left.-\psi\!\left(n_{xz}(i)+1\right)-\psi\!\left(n_{yz}(i)+1\right)\right].
\label{eq:frenzel-pompe-estimator}
\end{align}
We use the same standardization, maximum norm, and $k=3$ for MI and CMI, and likewise clip negative CMI estimates to zero.

For each of the $R$ realizations, we compute the corresponding MI or CMI estimate and obtain a distribution of estimates $\{\widehat{I}_r\}_{r=1}^R$.
We define the test statistic as the median of $\widehat{I}_r$ divided by its standard deviation:
\begin{equation}
   \widehat{T}=\frac{\operatorname{median}_r \widehat{I}_r}{\operatorname{std}_r \widehat{I}_r}.
\label{eq:posterior-integrated-test-statistic}
\end{equation}
This is the test statistic used to quantify the strength of correlation between different variables in a given gravitational-wave catalog.

\subsection{Event-permutation \texorpdfstring{$p$}{p}-value for catalog dependence}
To estimate the statistical significance of the observed correlation, we need a distribution of $\widehat{T}$ for the null hypothesis $H_0$.
We randomly permute the event labels of $Y$ relative to $X$ and recompute $\widehat{T}$ \cite{10.1093/biomet/asz024}.
This removes the event-to-event pairing and estimates the value of $\widehat{T}$ when there is no correlation between $X$ and $Y$.
We denote the statistic of the unpermuted catalog by $T_{\mathrm{obs}}$ and that of the $b$-th permuted catalog by $T_b$.
After $B$ permutations, we report the $p$-value as the fraction of $T_b$ values exceeding $T_{\mathrm{obs}}$.
$p=0$ indicates that none of the background permutations has an \ac{MI} value as strong as that of the observed catalog, but the resolution of the $p$-value is limited by the number of permutations $B$.
A small $p$ means that the observed pairing is unusual under the null hypothesis and provides evidence against independence.
However, to interpret a small $p$-value, we must consider that the independence null used here is not a null test of GR. Because the testing-GR and source parameters are inferred jointly from the same waveform, parameter degeneracies and/or source-dependent measurement precision can also induce catalog-level dependence even when GR is correct. Small permutation $p$-values therefore identify patterns that require further diagnosis, rather than evidence for a violation of GR.

We do not assign a permutation-based $p$-value to CMI because permuting $Y$ would necessarily destroy its relation with $Z$, breaking the conditional structure.
CMI is therefore reported only as a statistic without an associated significance estimate.
A valid null for CMI may require dedicated simulations of gravitational-wave events.

The analyses below use $B=1000$ unless otherwise stated. With 1000 permutations, A $p=0$ value should $p<10^{-3}$, since the permutation test has finite resolution.

\section{Validation with massive graviton injections}
\label{sec:massive-graviton-sim}

We use simulations to test whether the above framework recovers a known source-dependent waveform modification after event-level parameter estimation.
We inject a massive graviton corrected waveform and recover each signal with a parameterized post-Einstein (ppE) waveform rather than the underlying massive graviton waveform. The recovery parameter therefore differs from the physical parameter used in the injected theory.

\begin{figure*}[htbp]
\centering
\includegraphics[width=0.98\textwidth]{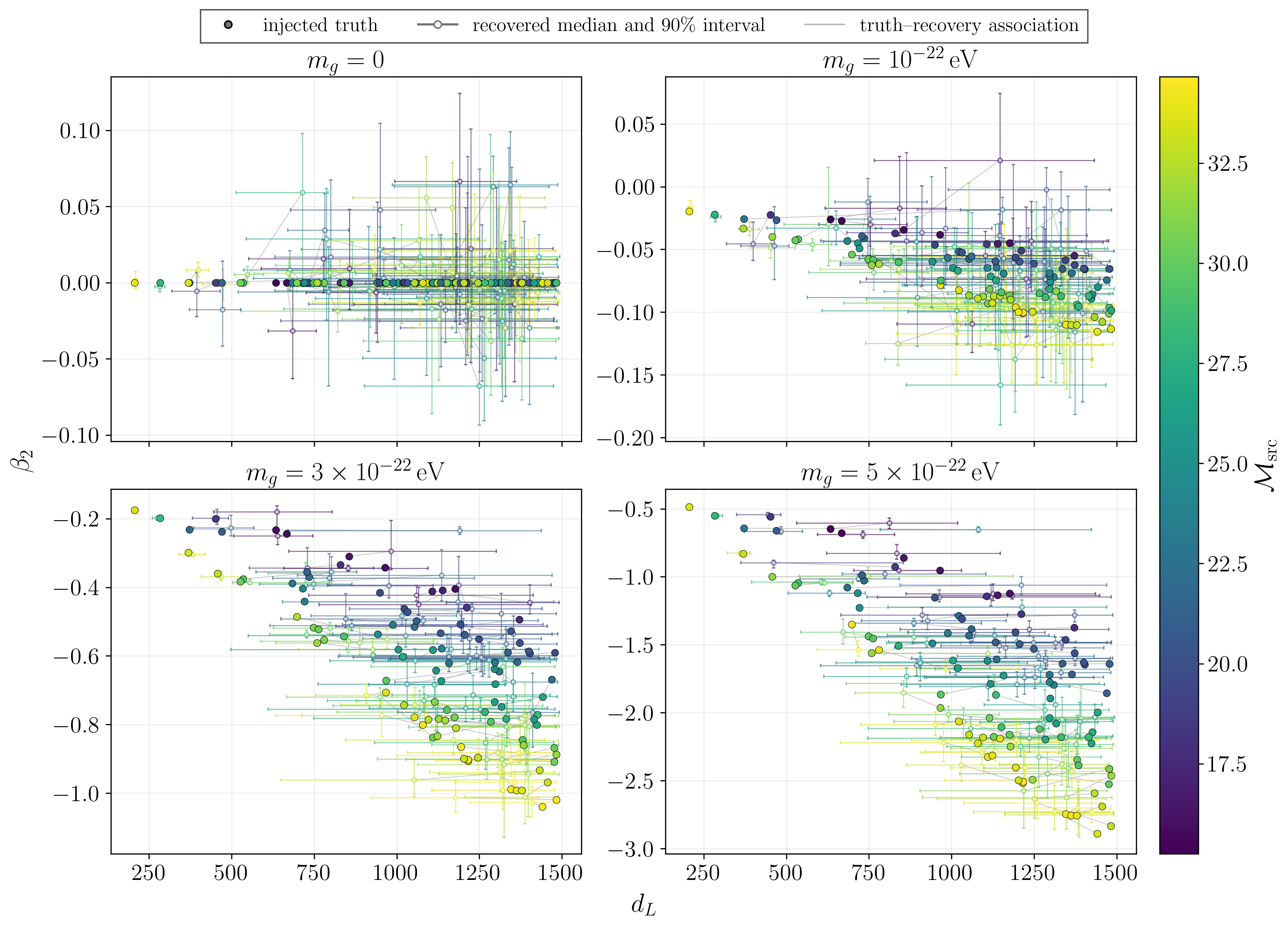}
\caption{Injected and recovered distance--$\beta_2$ relations for the four catalogs with $m_g=\{0,1,3,5\}\times10^{-22}\,\mathrm{eV}$, ordered from top left to bottom right. Filled circles mark injected true values, while open circles and horizontal and vertical error bars show recovered posterior medians and marginal 5th--95th percentiles. Gray segments associate each injected value with its recovered median. Color denotes the injected source-frame chirp mass.}
   \label{fig:mg-recovery-truth-association}
\end{figure*}

\begin{figure*}[t!]
\centering
\includegraphics[width=0.98\textwidth]{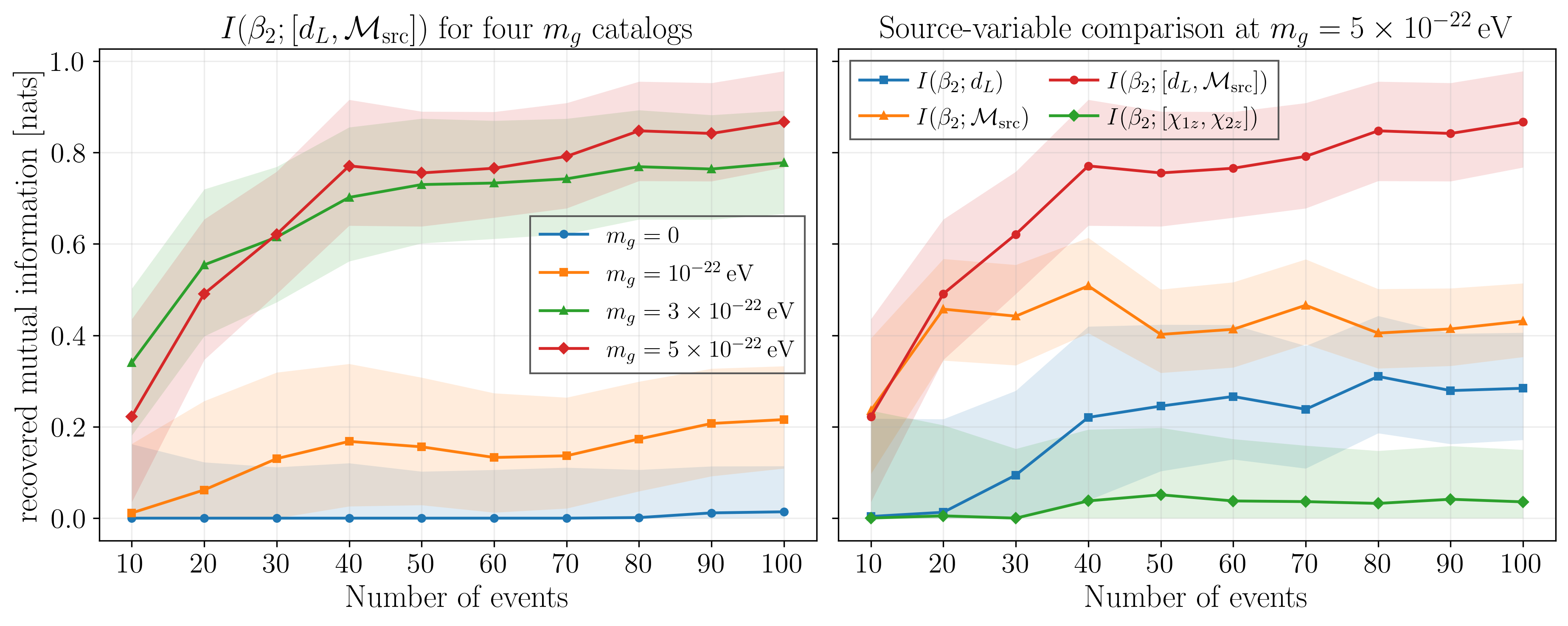}
\caption{Recovered MI as a function of the number of events. Left: joint distance--source chirp mass MI for all four massive graviton catalogs. Right: MI between $\beta_2$ and distance, source chirp mass, the joint of the two, and two spins, respectively, for $m_g=5\times10^{-22}\,\mathrm{eV}$. Curves show medians over 2000 posterior-resampled catalogs, and bands span the 5th--95th percentiles.}
   \label{fig:mg-fixed-prefix-recovered-mi}
\end{figure*}

\subsection{Mapping the massive graviton waveform to the ppE recovery parameter}

A nonzero graviton mass modifies the propagation speed of gravitational waves and produces a frequency-dependent phase correction~\cite{Will:1997bb,Mirshekari:2011yq,massiveg}.
The modified phase at frequency $f$ is
\begin{equation}
\Delta\Psi_{\rm MG}(f)=-\frac{\pi D_{\rm MG}(z)}{\lambda_g^2(1+z)f},
\label{eq:mg-phase-current}
\end{equation}
where $\lambda_g$ is the graviton Compton wavelength and
\begin{equation}
D_{\rm MG}(z)=\frac{1+z}{H_0}\int_0^z\frac{\mathrm{d}z'}{(1+z')^2E(z')},
\label{eq:mg-dc}
\end{equation}
where we use units with $c=1$. Here, $E(z)=H(z)/H_0$ where $H(z)$ is the Hubble expansion rate at redshift $z$, $H_0=H(0)$ is its present-day value.
The distance $D_{\rm MG}$ is the cosmological propagation factor entering the massive graviton phase, note that it is neither the luminosity distance nor the comoving distance.

We use the frequency-domain reduced-order implementation \texttt{SEOBNRv5HM\_ROM} of the \texttt{SEOBNRv5HM} waveform model~\cite{Pompili:2023tna} as the \ac{GR} baseline and modify it with the phase term in Eq.~\eqref{eq:mg-phase-current}.

The recovery waveform modifies the same \ac{GR} baseline with the parameterized post-Einstein phase term
\begin{equation}
\Delta\Psi_{\rm ppE}^{(2)}(f)=\frac{\beta_2}{\pi{\cal M}_{\rm det}f},
\label{eq:ppe-phase-current}
\end{equation}
where ${\cal M}_{\rm det}$ is the detector-frame chirp mass and our ppE convention labels this $f^{-1}$ coefficient as $\beta_2$, representing the deviation at first post-Newtonian (1PN) order.
Matching Eqs.~\eqref{eq:mg-phase-current} and \eqref{eq:ppe-phase-current} gives
\begin{equation}
\beta_2^{\rm MG}=-\pi{\cal M}_{\rm det}\left[\frac{\pi D_{\rm MG}(z)}{\lambda_g^2(1+z)}\right].
\label{eq:mg-ppebeta2-map}
\end{equation}
The detector- and source-frame chirp masses are related by ${\cal M}_{\rm det}=(1+z){\cal M}_{\rm src}$.
A common $m_g$ therefore maps to different effective values of $\beta_2$ for sources with different distances and chirp masses.
The recovery of $\beta_2$, rather than $m_g$, mimics a theory-agnostic analysis in which the physical origin of the deviation is not known a priori.

\subsection{Simulation and recovery}

We construct four simulation catalogs with $m_g=\{0,1,3,5\}\times10^{-22}\,\mathrm{eV}$, respectively.
Each simulation contains the same 100 source realizations and uses the same detector configuration and pseudo-random detector noise seeds, only the injected graviton mass changes.
The $m_g=0$ simulation reduces to the \ac{GR} case.

Source-frame chirp masses are drawn uniformly over $[15,35]\,M_\odot$. The mass ratio is defined as $q=m_1/m_2$ and drawn uniformly over $[1,3]$, while luminosity distances are drawn from a uniform-in-volume distribution over $[40,1500]\,\mathrm{Mpc}$.
Sky positions and orientations are isotropic, while the aligned spin components are drawn from a zero-mean Gaussian with standard deviation $0.2$, truncated to $[-0.5,0.5]$.
This binary black hole distribution is not necessarily the most realistic astrophysical population model consistent with what we know, but is representative enough to validate the \ac{MI} method.

Signals are generated with simulated LIGO Hanford, LIGO Livingston~\cite{LIGOScientific:2014pky}, and Virgo~\cite{VIRGO:2014yos} data at sensitivities typical of the fourth observing run (O4) \cite{LIGO:2024kkz,Capote:2024rmo}.
An injection is analyzed only when the optimal network \ac{SNR} exceeds 12, thereby selecting relatively loud events with reliable measurements.
Each event is recovered with the ppE waveform by sampling $\beta_2$ uniformly over $[-4,4]$. 
The other sampled parameters include source-frame chirp mass, mass ratio, aligned spins, luminosity distance, sky position, inclination, coalescence phase, and coalescence time.
The polarization is numerically marginalized in the likelihood calculation.
We use \texttt{PyCBC Inference}~\cite{Biwer:2018osg} for parameter estimation with the \texttt{dynesty} sampler~\cite{Speagle:2019ivv}, using 5000 live points and a stopping criterion of $\Delta\log Z=0.1$.

Figure~\ref{fig:mg-recovery-truth-association} presents the recovered $\beta_2$--$D_{\rm L}$ relation together with the injected values and source-frame chirp masses for all four catalogs.
Across the four catalogs, the injected true value of $\beta_2$ lies within the recovered central 90\% credible interval for 92, 91, 90, and 86 of 100 events as $m_g$ increases, indicating broad consistency between the injected and recovered values.

\subsection{Dependence test between the testing-\ac{GR} parameter and source properties}

To test whether the dependence remains detectable after parameter estimation, we estimate \ac{MI} for each simulated catalog.
The target quantities are the \ac{MI} values between the recovered $\beta_2$ and luminosity distance, source-frame chirp mass, and the joint of the two.

Figure~\ref{fig:mg-fixed-prefix-recovered-mi} evaluates \ac{MI} of the paired variables as a function of accumulated event number from 10 to 100 events in steps of 10.
For the MI between $\beta_2$ and the joint source-frame chirp mass and luminosity distance, the full 100-event catalogs give $T_{\rm obs}=0.353$, $3.134$, $11.425$, and $13.679$ for $m_g=\{0,1,3,5\}\times10^{-22}\,\mathrm{eV}$, respectively.
The statistical meaning of these values is that the median MI lies $11.425$ and $13.679$ standard deviations above zero for the two largest graviton masses.
As expected, the $m_g=3\times10^{-22}\,\mathrm{eV}$ and $m_g=5\times10^{-22}\,\mathrm{eV}$ catalogs show a rising trend as the number of events increases.
The $m_g=1\times10^{-22}\,\mathrm{eV}$ catalog shows a weak rising trend.
The controlled \ac{GR} simulation shows no correlation, with MI indistinguishable from zero.

In the right panel of Fig.~\ref{fig:mg-fixed-prefix-recovered-mi}, we focus on the $m_g=5\times10^{-22}\,\mathrm{eV}$ catalog and compare the MI values for various pairs of variables.
The MI between $\beta_2$ and the joint distance and mass variable is larger than the MI between $\beta_2$ and either one, indicating that both source-frame chirp mass and luminosity distance contribute to the recovered correlation.
The nonmonotonic dependence on one-dimensional mass as the number of events increases may be due to random fluctuations in the event ordering with mass and distance. Nevertheless, the MI remains distinguishable from zero.
We deliberately include the MI between $\beta_2$ and the two spin magnitudes as a test.
The result shows no correlation with the testing-\ac{GR} parameter as expected.

In our statistical framework, $T_{\rm obs}$ quantifies the strength of correlation between the testing-\ac{GR} parameter and source properties.
We assess the significance of the joint correlation $I(\beta_2;[D_L,{\cal M}_{\rm src}])$ by comparing $T_{\rm obs}$ with the event-permutation null distribution.
For $m_g=\{0,1,3,5\}\times10^{-22}\,\mathrm{eV}$, the corresponding exceedance counts are $48/1000$, $0/1000$, $0/1000$, and $0/1000$, giving $p=0.048$, $0$, $0$, and $0$, respectively.
The $m_g=0$ GR catalog shows a relatively small $p$-value. This likely arises because distant events have broader posteriors, which can induce a weak correlation between  $\beta_2$ and distance even when $m_g=0$.
In such cases, an end-to-end simulation with a realistic population model and selection effects would be required to obtain the true null distribution, rather than relying solely on permutation tests.
Nevertheless, the median MI is only $T_{\rm obs}=0.353$ standard deviations above zero, indicating that the detected dependence is weak and consistent with zero within the posterior-resampling uncertainty.
We therefore conclude that our framework is sufficiently sensitive to recover the known correlation in the $m_g>0$ catalogs, while showing no significant evidence of correlation in the \ac{GR} catalog.

Since the injected $\beta_2$ is a function of both distance and source mass, we also compute the CMI between $\beta_2$ and each variable conditioned on the other.
The result is reported in Table~\ref{tab:mg-cmi-placeholder}.
After conditioning on source mass, the residual distance dependence remains distinguishable from zero, and the same is true for the chirp-mass dependence conditioned on distance.
Together with the fact that the joint MI is larger than either one, this indicates that both source mass and distance contribute to the recovered correlation.

\begin{table}[t]
\caption{CMI statistics. Each row reports the posterior-resampled CMI median divided by its standard deviation.}
\label{tab:mg-cmi-placeholder}
\begin{ruledtabular}
\begin{tabular}{ccc}
$m_g/(10^{-22}\,\mathrm{eV})$ & CMI diagnostic & $T_{\rm CMI}$ \\
\hline
$0$ & $I(\beta_2;D_L\mid {\cal M}_{\rm src})$ & 0.400 \\
$0$ & $I(\beta_2;{\cal M}_{\rm src}\mid D_L)$ & 0.052 \\
$1$ & $I(\beta_2;D_L\mid {\cal M}_{\rm src})$ & 2.180 \\
$1$ & $I(\beta_2;{\cal M}_{\rm src}\mid D_L)$ & 2.162 \\
$3$ & $I(\beta_2;D_L\mid {\cal M}_{\rm src})$ & 5.976 \\
$3$ & $I(\beta_2;{\cal M}_{\rm src}\mid D_L)$ & 8.929 \\
$5$ & $I(\beta_2;D_L\mid {\cal M}_{\rm src})$ & 6.617 \\
$5$ & $I(\beta_2;{\cal M}_{\rm src}\mid D_L)$ & 11.533 \\
\end{tabular}
\end{ruledtabular}
\end{table}

\section{Real-data testing-\ac{GR} results}
\label{sec:real-data}

We apply the dependence test to four public event-level testing-\ac{GR} products: the Test Infrastructure for General Relativity (TIGER), the flexible theory-independent (FTI) test, modified dispersion relations (MDR), and parity violation.
The parity-violation analysis uses posterior samples released with Ref.~\cite{Wang:2021birefringence4OGC}.
The other analyses use samples released with the GWTC-4.0 testing-\ac{GR} study~\cite{LIGOScientific:2026fcf,ligo_scientific_collaboration_2026_21403342}.

For each analysis, we first summarize how the testing-\ac{GR} parameter enters the waveform.
We then evaluate its dependence on the available source and measurement properties, including source frame chirp mass $\mathcal{M}^{\rm src}$, total mass $M^\mathrm{src}$, mass ratio $q$, effective spin $\chi_{\rm eff}$ and precessing spin $\chi_{\rm p}$, luminosity distance $D_L$, inclination angle $\theta_\mathrm{JN}$, sky location, and \ac{SNR} $\rho_\mathrm{net}$.
Finally, we inspect the event-level posteriors for the largest value of $T_{\rm obs}$.
These correlation studies serve as exploratory diagnostics of structure in the released posteriors.

We observe some limitations in identifying whether the dependence is physical or produced by parameter degeneracy within the waveform model using $T_{\rm obs}$ and $p$-values.
To investigate, we therefore also resample the two variables independently within every event and recompute the test statistics for the strongest pair in each analysis.
This procedure preserves their event-level marginal posteriors but removes their within-event joint-posterior coupling.
A large decrease in $T_{\rm obs}$ indicates that parameter degeneracy dominates the result.
If the statistic persists, the dependence is instead carried mainly by the catalog level correlation.

\subsection{Testing Infrastructure for General Relativity}

TIGER introduces fractional deviations into the coefficients of an inspiral--merger--ringdown phenomenological waveform~\cite{Li:2011cg,Agathos:2013upa,Meidam:2017dgf,Roy:2025gzv, LIGOScientific:2026fcf}.
In the inspiral phase, only the nonspinning part of each post-Newtonian coefficient is deformed,
\begin{equation}
\varphi_i \longrightarrow
 (1+\delta\hat\varphi_i)\varphi_i^{\rm GR,NS}
 +\varphi_i^{\rm GR,S},
\end{equation}
where $\varphi_i^{\rm GR,NS}$ and $\varphi_i^{\rm GR,S}$ denote the nonspinning and spinning contributions at post-Newtonian order $i$.
Analogous fractional deformations are applied to the intermediate coefficients $b_j$ and merger--ringdown coefficients $c_j$, with
\begin{equation}
c_j \longrightarrow (1+\delta\hat c_j)c_j^{\rm GR},
\label{eq:tiger-parameterization}
\end{equation}
where $c_j$ denotes a merger--ringdown calibration coefficient.

The GWTC-4.0 release contains joint deviation--source posteriors for 18 testing-\ac{GR} parameters, including $\delta\hat\varphi_i$, $\delta\hat b_j$, and $\delta\hat c_j$.
The event set differs among tests because each waveform segment is informative for a different subset of signals.
The inspiral-coefficient tests contain 52 events, except $\delta\hat\varphi_0$, which contains 50.
The post-inspiral tests contain 68 events.

\begin{figure*}[p!]
   \centering
   \includegraphics[width=\textwidth]{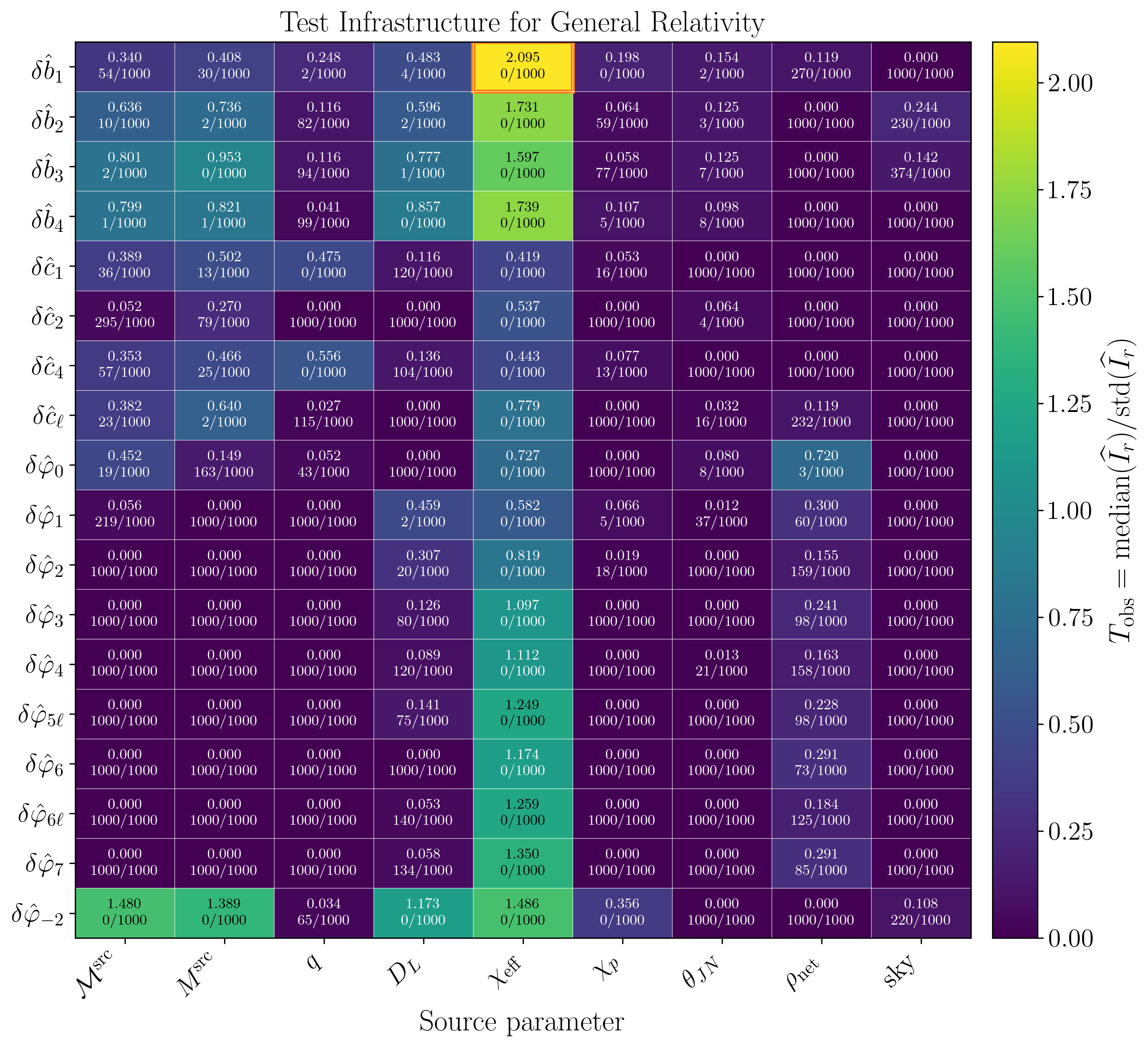}
   \caption{Dependence map for TIGER parameters and source or measurement properties. Color and the upper cell entry give $T_\mathrm{obs}$; the lower entry gives the permutation-based $p$-value. The post-inspiral rows ($\delta\hat b$ and $\delta\hat c$) use 68 events. The inspiral rows ($\delta \hat\varphi$) use 52 events, except $\delta\hat\varphi_0$, which uses 50. The orange cell marks the largest value.}
   \label{fig:realdata-tiger}
\end{figure*}

\begin{figure}[H]
   \centering
   \includegraphics[width=\columnwidth]{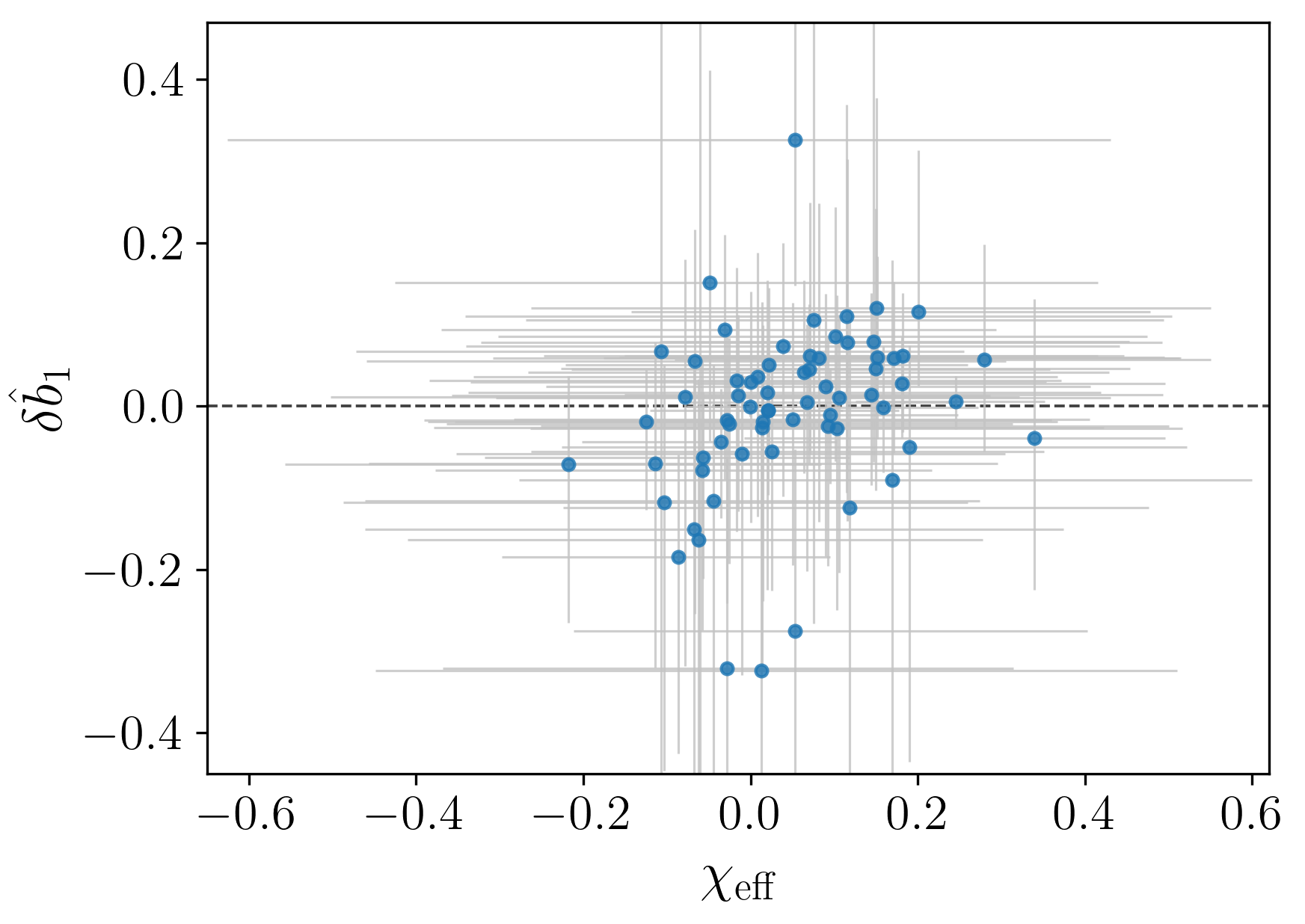}
   \caption{Event-level posterior for the largest TIGER cell, $\delta\hat b_1$ versus $\chi_{\rm eff}$, across 68 events. Here, $T_{\rm obs}=2.095$ and the permutation-based $p$-value is $0$. Points and bars show marginal posterior medians and 5th--95th percentiles. All events and their full joint posteriors are used in the MI calculation. Independent non-joint within-event resampling reduces $T_{\rm obs}$ from $2.095$ to $0.271$, indicating that within-event joint-posterior coupling dominates the dependence.}
   \label{fig:realdata-tiger-events}
\end{figure}

Figure~\ref{fig:realdata-tiger} summarizes the dependence of these parameters on nine source and measurement properties.
The largest cell is $\delta\hat b_1$ versus $\chi_{\rm eff}$, with $T_\mathrm{obs}=2.095$ and a permutation-based $p=0$.
The $\delta\hat b_{2,3,4}$ rows show similar, but weaker, dependence on $\chi_{\rm eff}$.
Figure~\ref{fig:realdata-tiger-events} shows the event-level marginal posteriors for $\delta\hat b_1$ and $\chi_{\rm eff}$ across 68 events.
Their medians exhibit mild positive ordering.
An independent resampling for each variable reduces $T_{\rm obs}$ from $2.095$ to $0.271$.
Hence the dependence is dominated by within-event joint-posterior coupling, which may be due to a $\delta\hat b_1$--$\chi_{\rm eff}$ parameter degeneracy or prior-induced coupling.

In addition, the negative order post-Newtonian parameter $\delta\hat\varphi_{-2}$ is mildly associated with the source-frame mass, with $T_\mathrm{obs}=1.480$.
It is qualitatively consistent with the known covariance between negative post-Newtonian phase terms and chirp mass~\cite{Sanger:2024axs}.

\subsection{Flexible theory-independent test}

\begin{figure*}[t]
   \centering
   \includegraphics[width=0.98\textwidth]{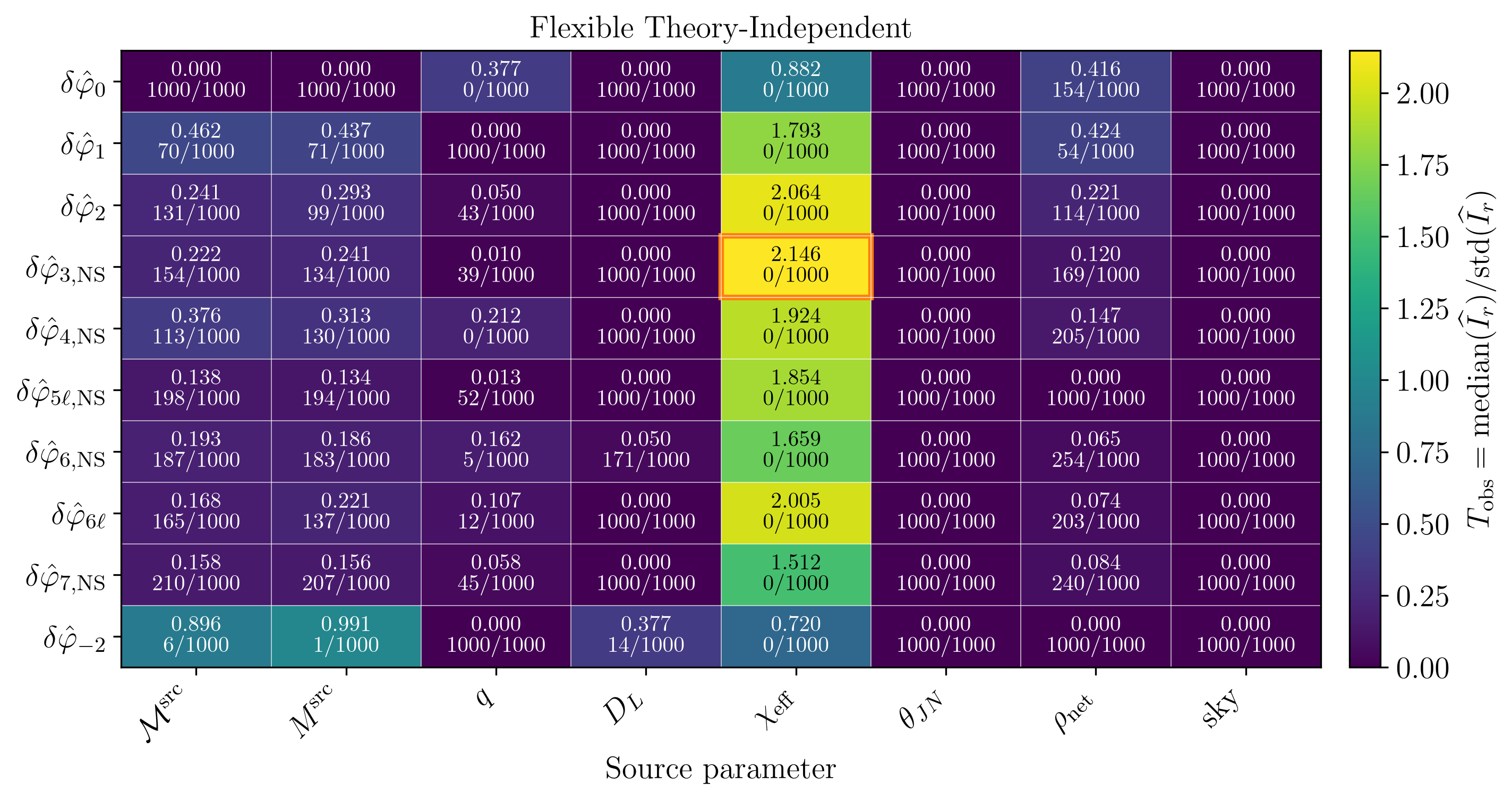}
   \caption{Dependence map for FTI parameters and source or measurement properties. Color and the upper cell entry give $T_\mathrm{obs}$; the lower entry gives the permutation-based $p$-value. The $\delta\hat\varphi_0$ row uses 14 events, whereas every other row uses 18. The orange cell marks the largest value.}
   \label{fig:realdata-fti}
\end{figure*}

The FTI analysis modifies the post-Newtonian inspiral phase by adding a correction to the baseline waveform and tapering it to zero before merger~\cite{Mehta:2022pcn,LIGOScientific:2026fcf}.
Schematically,
\begin{equation}
\Psi_{\rm FTI}(f)=\Psi_{\rm GR}(f) +W(f)\sum_i\delta\hat\varphi_i B_i(f),
\label{eq:fti-parameterization}
\end{equation}
where $B_i(f)$ gives the frequency dependence at post-Newtonian order $i$, and $W(f)$ is the inspiral taper.
The GWTC-4.0 release supplies joint posteriors for 10 deviation parameters at different post-Newtonian orders.
Most tests contain 18 events, whereas the $\delta\hat\varphi_0$ test contains 14.
The FTI analysis uses an aligned-spin waveform, so the precession parameter $\chi_p$ is not included.
The MI results are shown in Fig.~\ref{fig:realdata-fti}.

The dominant feature of Fig.~\ref{fig:realdata-fti} is the repeated dependence on $\chi_{\rm eff}$ across several post-Newtonian orders.
The largest pair is $\delta\hat\varphi_{3,\mathrm{NS}}$ versus $\chi_{\rm eff}$, with $T_{\rm obs}=2.146$ and a permutation-based $p=0$.

\begin{figure}[t]
   \centering
   \includegraphics[width=\columnwidth]{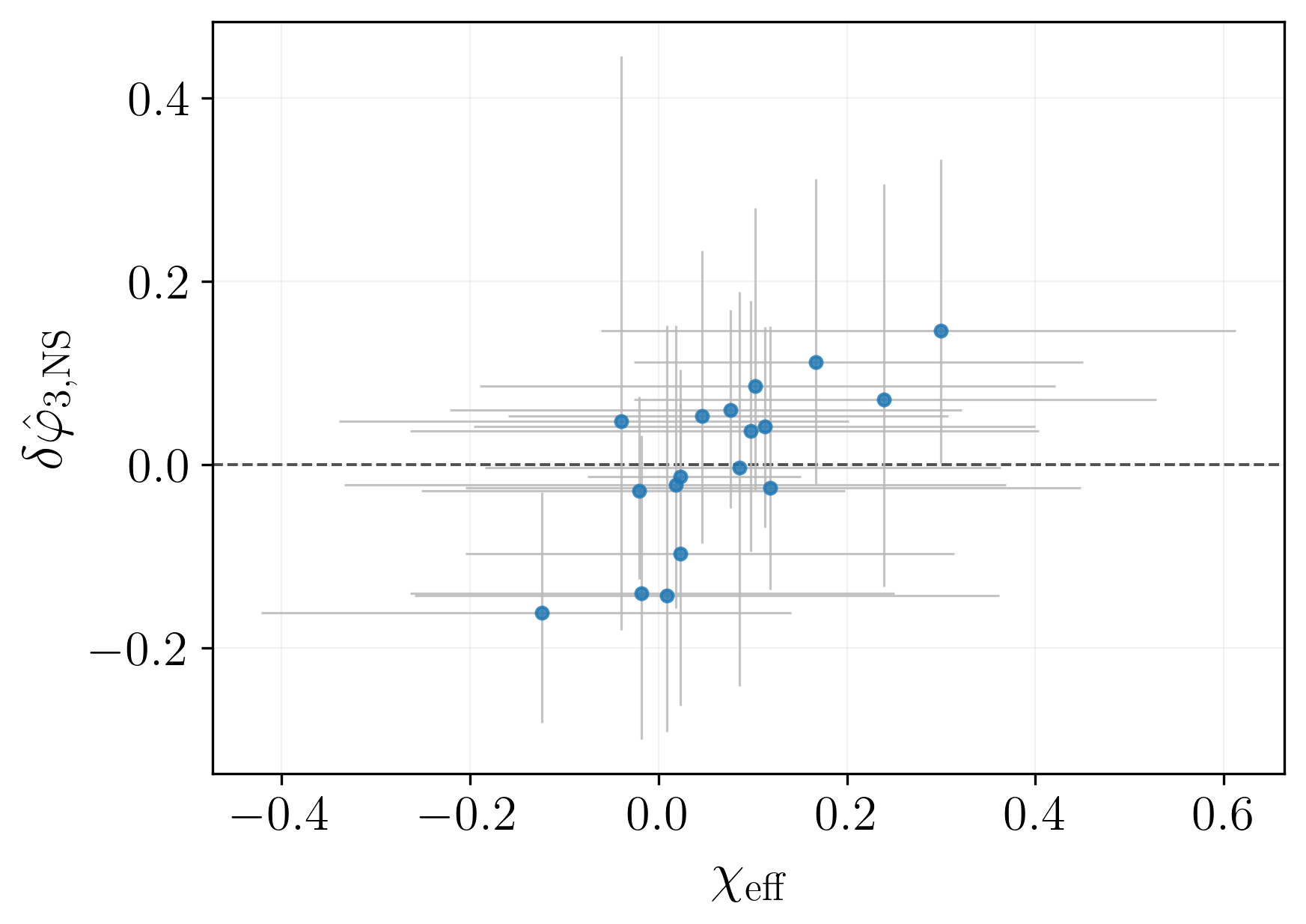}
   \caption{Event-level posterior summaries for the largest FTI cell, $\delta\hat\varphi_{3,\mathrm{NS}}$ versus $\chi_{\rm eff}$, across 18 events. Here, $T_{\rm obs}=2.146$ and the permutation-based $p$-value is $0$. Points and bars show marginal posterior medians and 5th--95th percentiles. Independent within-event resampling reduces $T_{\rm obs}$ from $2.146$ to $0.187$, indicating that within-event joint-posterior coupling dominates the dependence.}
   \label{fig:realdata-fti-events}
\end{figure}

Figure~\ref{fig:realdata-fti-events} shows mild positive ordering of the marginal medians of the FTI parameters with $\chi_{\rm eff}$, similar to \cref{fig:realdata-tiger-events}, but this trend does not persist without the within-event covariance.
Independent within-event resampling reduces $T_{\rm obs}$ from $2.146$ to $0.187$.
As in the TIGER result, this result motivates a spin--testing-GR parameter degeneracy interpretation.

\subsection{Modified dispersion relation test}
The MDR test modifies gravitational-wave propagation through a phenomenological dispersion relation~\cite{Baka:2025drk,LIGOScientific:2026fcf},
\begin{equation}
E^2 = p^2c^2+A_\alpha p^\alpha c^\alpha,
\end{equation}
where $E$ is the graviton energy, $p$ is its momentum, $A_\alpha$ is the dispersion coefficient, and $\alpha$ specifies the power-law modification.
The corresponding phase modification is
\begin{equation}
\delta\Psi_\alpha(f) \propto A_\alpha D_\alpha(z)f^{\alpha-1}.
\label{eq:mdr-parameterization}
\end{equation}
Here, $f$ is the detector-frame gravitational-wave frequency and $D_\alpha(z)$ is the $\alpha$-dependent cosmological propagation distance factor for a source at redshift $z$ \cite{LIGOScientific:2026fcf}.
The case $\alpha=0$ corresponds to a massive graviton, with $A_0\propto m_g^2$.

\begin{figure*}[t]
   \centering
   \includegraphics[width=0.98\textwidth]{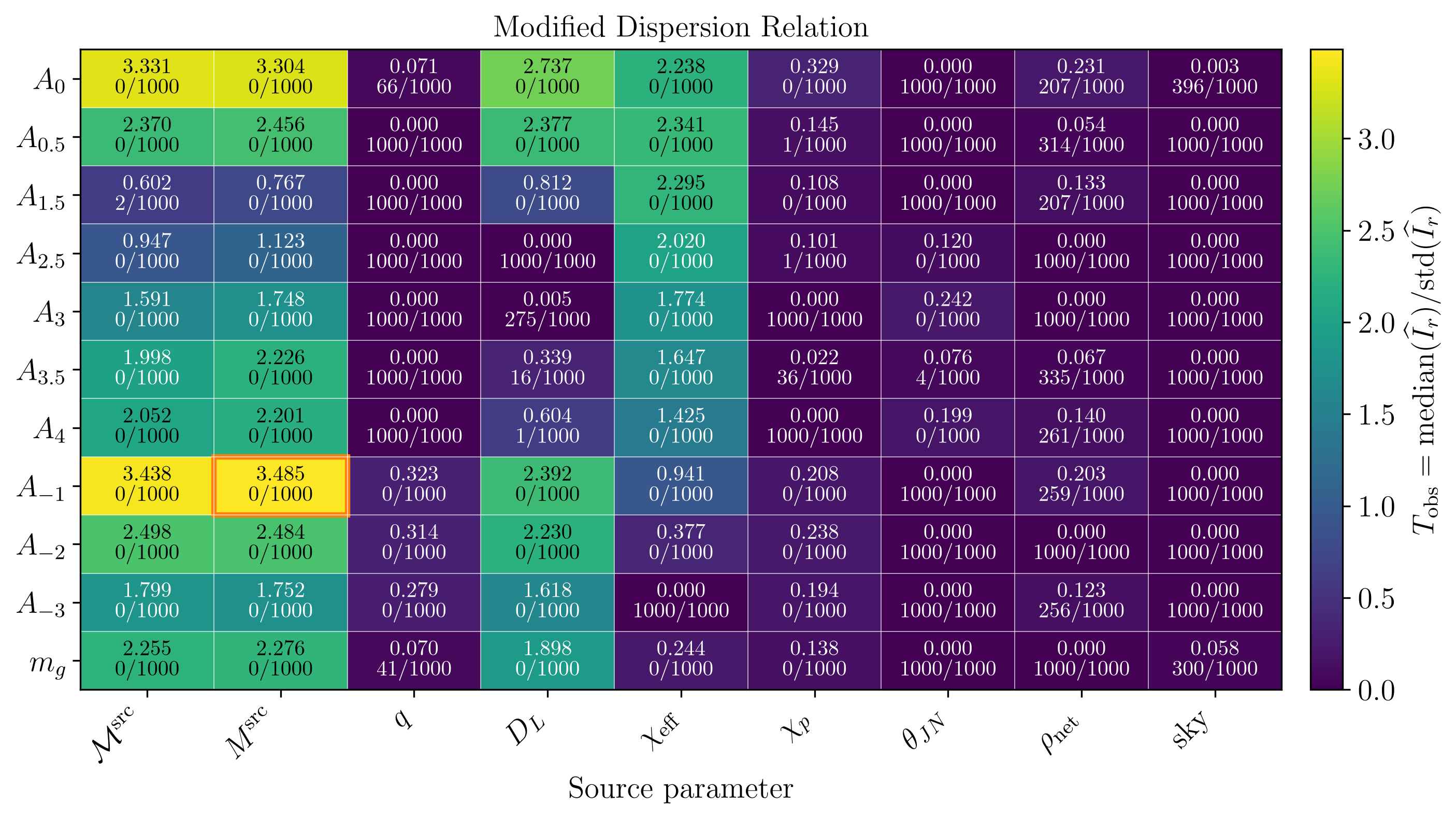}
   \caption{Dependence map for MDR parameters and source or measurement properties using joint posteriors. Color and the upper cell entry give $T_\mathrm{obs}$; the lower entry gives the permutation-based $p$-value. Every row uses 84 events, and the orange cell marks the largest value.}
   \label{fig:realdata-mdr}
\end{figure*}

The GWTC-4.0 release provides joint posteriors for 10 MDR coefficients and $m_g$ across 84 events.
Figure~\ref{fig:realdata-mdr} shows repeated dependence on source-frame mass and on distance or redshift across several rows.
The largest cell is $A_{-1}$ versus source-frame total mass, with $T_{\rm obs}=3.485$ and a permutation-based $p=0$.
The massive graviton row also depends on mass, with $T_{\rm obs}=2.255$ for chirp mass and $2.276$ for total mass.
Both cells have permutation-based $p=0$.

\begin{figure}[htbp]
   \centering
   \includegraphics[width=\columnwidth]{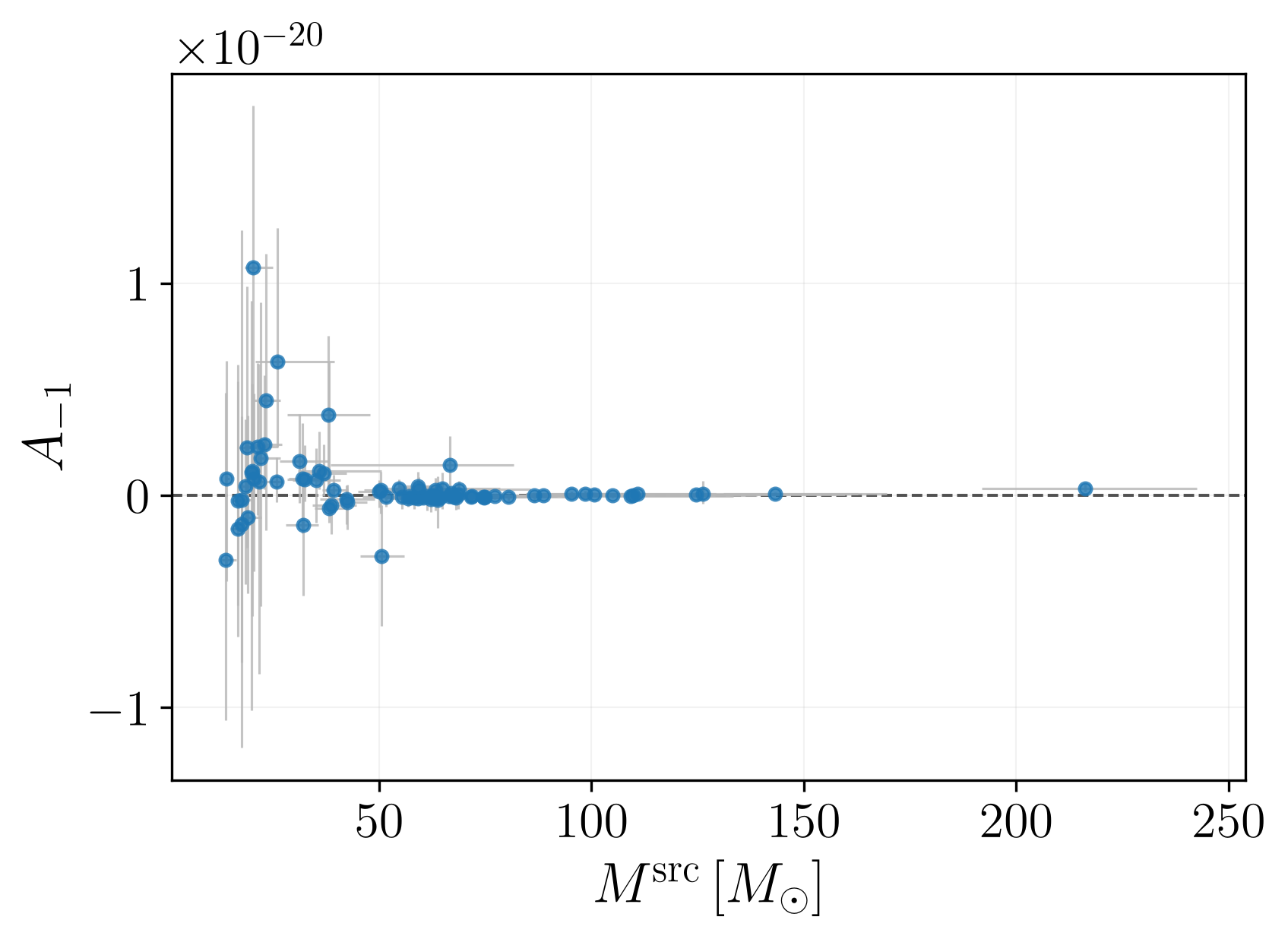}
   \caption{Event-level posterior summaries for the largest MDR cell, $A_{-1}$ versus source-frame total mass, across 84 events. Here, $T_{\rm obs}=3.485$ and the permutation-based $p$-value is $0$. Points and bars show marginal posterior medians and 5th--95th percentiles. Independent within-event resampling changes $T_{\rm obs}$ from $3.485$ to $3.720$, showing that direct within-event joint-posterior coupling is not the primary origin.}
   \label{fig:realdata-mdr-events}
\end{figure}

Figure~\ref{fig:realdata-mdr-events} examines the posterior of $A_{-1}$ and $M^\mathrm{src}$.
The medians of $A_{-1}$ remain concentrated near the \ac{GR} value over most of the mass range but the error bar narrows as mass increases.
The dependence therefore reflects source-dependent constraining power rather than a common nonzero $A_{-1}$ that varies with mass.
This nonlinear change in posterior width is detectable with MI but may be missed by a linear correlation coefficient.

Independent within-event resampling changes $T_{\rm obs}$ from $3.485$ to $3.720$, so unlike the TIGER and FTI results, the dependence is not primarily driven by direct within-event parameter coupling.
However, the test does not exclude waveform degeneracies or prior effects that alter the marginal posterior widths as a function of mass.

\subsection{Parity violation}

The parity-violation analysis allows left- and right-handed circular polarizations to acquire opposite propagation related modifications~\cite{Wang:2020cub,Wang:2021birefringence4OGC},
\begin{equation}
\tilde h_{L,R}^{\rm PV}(f)
 =\tilde h_{L,R}^{\rm GR}(f)
 \exp\!\left[\mp i\,\delta\Psi\!\left(f;M_{\rm PV}^{-1}\right)\right]
\end{equation}
where the phase correction is
\begin{equation}
 \delta\Psi \propto M_{\rm PV}^{-1}f^2.
\label{eq:parity-parameterization}
\end{equation}
$M_{\rm PV}^{-1}$ is the signed inverse parity-violation energy scale in the effective field theory framework.

\begin{figure}[t]
   \centering
   \includegraphics[width=\columnwidth]{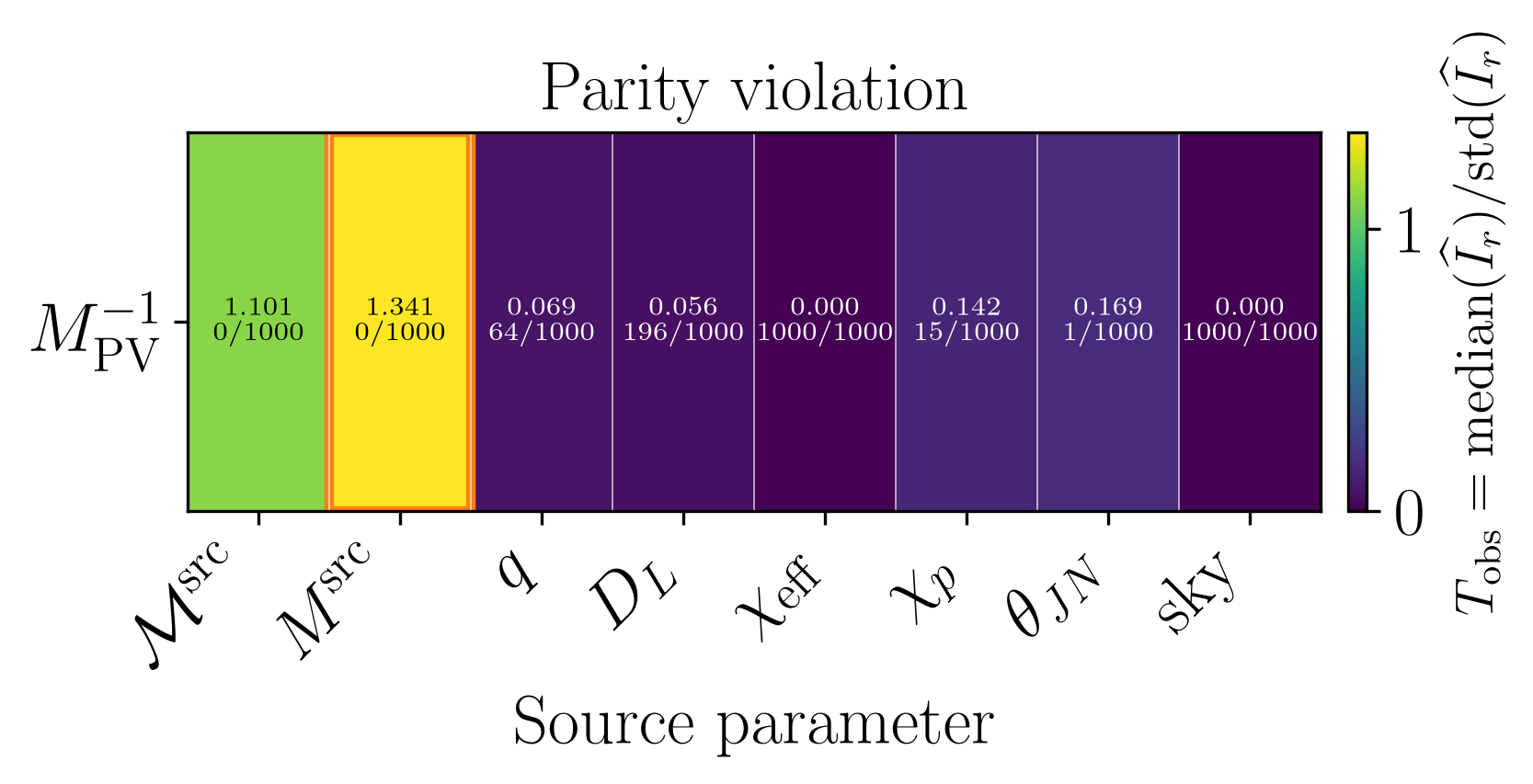}
   \caption{Parity-violation dependence map for the $M_{\rm PV}^{-1}$ and source or measurement properties. Color and the upper cell entry give $T_\mathrm{obs}$; the lower entry gives the permutation-based $p$-value. The row uses 94 events for non-sky variables and 93 for sky location (without GW170817 which fixes the sky location). The orange cell marks the largest value.}
   \label{fig:realdata-parity}
\end{figure}

We analyze $M_{\rm PV}^{-1}$ using the 94-event 4-OGC release~\cite{Nitz:2021zwj,Wang:2021birefringence4OGC}.
The result is shown in Fig.~\ref{fig:realdata-parity}.
The largest association is from source-frame total mass, with $T_{\rm obs}=1.341$ and a permutation-based $p=0$.
Source-frame chirp mass follows with $T_{\rm obs}=1.101$ and $p=0$.

\begin{figure}[t!]
   \centering
   \includegraphics[width=\columnwidth]{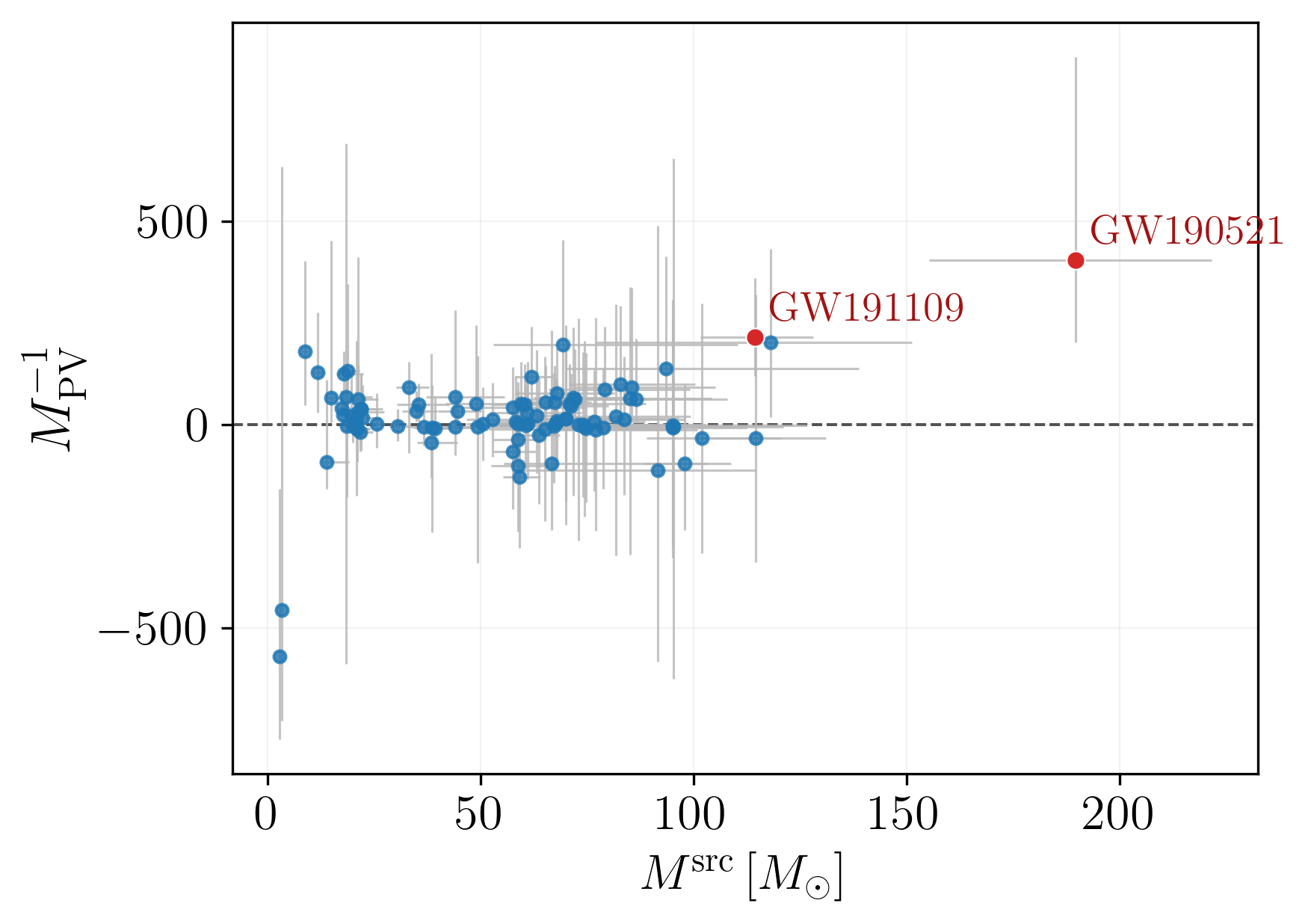}
   \caption{Event-level posterior summaries for $M_{\rm PV}^{-1}$ versus source-frame total mass across 94 events. Here, $T_{\rm obs}=1.341$ and the permutation-based $p$-value is $0$. Points and bars show marginal posterior medians and 5th--95th percentiles, with GW190521 and GW191109 highlighted. Independent within-event resampling changes $T_{\rm obs}$ from $1.341$ to $1.153$, showing that most of the dependence remains after the joint-posterior pairing is removed.}
   \label{fig:realdata-parity-events}
\end{figure}

As pointed out by \cite{Wang:2021birefringence4OGC}, GW190521 and GW191109\_010717 (hereafter GW191109)'s posteriors are the most inconsistent with zero among the binary black hole events and give the largest deviations in $M_{\rm PV}^{-1}$.
Their $M_{\rm PV}^{-1}$ posterior medians and intervals are marked in red in Fig.~\ref{fig:realdata-parity-events}.
Excluding these two events reduces the total-mass statistic from $1.341$ to $1.118$.
They therefore strengthen, but do not fully determine, the association.
Independent within-event resampling changes $T_{\rm obs}$ from $1.341$ to $1.153$, therefore most of the dependence survives removal of the joint-posterior pairing, similar to the MDR results and unlike the TIGER and FTI results.

\section{Discussion and conclusions}
\label{sec:discussion-conclusion}

We introduced a catalog-level test for dependence between phenomenological testing-\ac{GR} parameters and source or measurement properties.
This question complements event-level GR tests and hierarchical analyses of the deviation-parameter population.
Posterior resampling propagates event-level measurement uncertainty, while \ac{MI} captures dependence without assuming a functional form.
Event permutations then test whether the observed pairing is unusual under a source-independence null.
To diagnose whether the dependence is driven by parameter degeneracy, we independently resample the two parameters within each event, thereby removing their joint-posterior pairing while preserving their marginal posteriors.
This framework is intended to diagnose the origin of apparent deviations by identifying systematic trends, not to provide standalone evidence against \ac{GR}.

The massive graviton injections demonstrate that the method recovers a known source-dependent modification even when the recovery model is theory agnostic.
The propagation phase maps onto a ppE parameter that depends jointly on source-frame chirp mass and luminosity distance.
For $m_g=\{0,1,3,5\}\times10^{-22}\,\mathrm{eV}$, the recovered joint dependence gives $T_{\rm obs}=0.353$, $3.134$, $11.425$, and $13.679$, respectively.
The two largest injections are therefore recovered clearly, while the \ac{GR} catalog remains consistent with zero dependence within posterior-resampling uncertainty.

The population level resampled posteriors contain two conceptually different sources of dependence.
The first is event-to-event variation in the marginal posteriors, which can reflect a real physical dependence, measurement precision, or selection effects.
The second is within-event joint-posterior degeneracy, which can arise from a parameter degeneracy within the waveform parameterization or from the prior.
We separate these contributions by sampling from the marginals while removing their pairing.

This diagnostic produces a clear contrast among the four real-data analyses.
For TIGER, $T_{\rm obs}$ decreases from $2.095$ to $0.271$ after the within-event pairing is removed.
For FTI, it decreases from $2.146$ to $0.187$.
These reductions show that the strongest TIGER and FTI cells are dominated by within-event parameter degeneracies.

The MDR and parity-violation results behave differently.
For MDR, the event-level posteriors of $A_{-1}$ narrow systematically toward higher source-frame total mass.
The dependence is therefore carried mainly by mass-dependent measurement precision.
For parity violation, most of the association also survives.
Its main visible structure is associated with the high mass outliers GW190521 and GW191109, whose posterior medians are displaced from zero.
The exact physical origin of the dependence thus needs further investigation.

Several limitations remain for our methodology.
The present analysis gives all events equal weight, which can lead to biased results. Some regions of source-parameter space may be more likely to produce broad or apparently nonzero testing-\ac{GR} posteriors, which needs to be addressed.
The nearest-neighbour estimators also depend on catalog size, posterior-sample count, parameter dimension, preprocessing, and the choice of $k$.
Simulated catalogs with known dependence may be needed to quantify the systematic uncertainty associated with these choices.

The next step is to investigate the origin of selected dependence using targeted reanalyses.
One can perform leave-one-out tests to identify whether certain events are dominant the dependence, while alternative waveforms and priors can test parameterization dependence.
Injection and recovery studies can determine how often similar structures arise under \ac{GR}.
Data-quality checks and explicit selection models are also helpful before assigning a physical interpretation.
The framework can be extended to more testing-GR tests such as ringdown, remnant, and inspiral--post-merger consistency tests.
Additional source properties can include eccentricity, precession, tidal deformability, and quantitative measures of data quality.
When a genuine dependence is detected, regression methods can reconstruct its functional form and can improve detection sensitivity if a suitable parameterized test is performed.
This framework therefore provides a systematic path from detecting catalog-level dependence to investigating whether its origin is physical or associated with gravitational-wave inference systematics.

\begin{acknowledgments}
%I thank Bruce Allen for suggesting that I write single-author papers during a postdoctoral annual review at AEI Hannover, which motivated this paper.
YFW thanks Xikai Shan, Soumen Roy, and Suvodip Mukherjee for the comments on this manuscript and Elise Sänger for the comments on a presentation of this work. YFW also thanks Bruce Allen for the suggestion to write single author papers, which motivates this work.
The computational work for this manuscript was carried out on the Hypatia computer cluster at the Max Planck Institute for Gravitational Physics (Albert Einstein Institute) in Potsdam.
This research has made use of data or software obtained from the Gravitational Wave Open Science Center (gwosc.org), a service of the LIGO Scientific Collaboration, the Virgo Collaboration, and KAGRA. This material is based upon work supported by NSF's LIGO Laboratory which is a major facility fully funded by the National Science Foundation, as well as the Science and Technology Facilities Council (STFC) of the United Kingdom, the Max-Planck-Society (MPS), and the State of Niedersachsen/Germany for support of the construction of Advanced LIGO and construction and operation of the GEO600 detector. Additional support for Advanced LIGO was provided by the Australian Research Council. Virgo is funded, through the European Gravitational Observatory (EGO), by the French Centre National de Recherche Scientifique (CNRS), the Italian Istituto Nazionale di Fisica Nucleare (INFN) and the Dutch Nikhef, with contributions by institutions from Belgium, Germany, Greece, Hungary, Ireland, Japan, Monaco, Poland, Portugal, Spain. KAGRA is supported by Ministry of Education, Culture, Sports, Science and Technology (MEXT), Japan Society for the Promotion of Science (JSPS) in Japan; National Research Foundation (NRF) and Ministry of Science and ICT (MSIT) in Korea; Academia Sinica (AS) and National Science and Technology Council (NSTC) in Taiwan.
\end{acknowledgments}
%%%%%%%%%%%%%%%%%%%%%%%%%%%%%%%%%%

\bibliography{refer}
\end{document}